\documentclass[12pt,preprint]{elsarticle}
\PassOptionsToPackage{hyphens}{url}
\usepackage{lineno,hyperref}

\modulolinenumbers[5]
\usepackage[utf8]{inputenc}
\makeatletter
\def\ps@pprintTitle{%
 \let\@oddhead\@empty
 \let\@evenhead\@empty
 \def\@oddfoot{}%
 \let\@evenfoot\@oddfoot}   
\makeatother
\usepackage{subfig}  
\usepackage{titlesec}

\titleformat{\paragraph}
{\normalfont\normalsize\itshape}{\theparagraph}{1em}{}
\titlespacing*{\paragraph}
{0pt}{3.25ex plus 1ex minus .2ex}{0.2ex plus 1.2ex}

\usepackage{float}
\usepackage{amsmath,bm}  
\usepackage{longtable}
\usepackage[ruled,vlined]{algorithm2e}
\SetKwComment{Comment}{$\triangleright$\ }{}
\usepackage[noend]{algpseudocode}
\usepackage{pdflscape}
\usepackage{tabularx} 
\usepackage[usestackEOL]{stackengine}
\usepackage{booktabs} 
\usepackage[table]{xcolor} 
\usepackage{amsfonts} 
\usepackage{natbib}
\usepackage{multirow}
\usepackage{array}
\begin{document}
\begin{frontmatter}
\title{Sequence to sequence AE-ConvLSTM network for modelling the dynamics of PDE systems}

\author{Priyesh Rajesh Kakka$^{1}$,  \\
        \small $^{1}$pkakka@nd.edu, \\
        \small $^{1}$University of Notre Dame}
      
\date{August 2021}

\begin{abstract}
The article explains the convolutional LSTM (ConvLSTM) network~\cite{Shi2015CLSTM} in detail and introduces an improved auto-encoder version of the ConvLSTM network called AE-ConvLSTM. AE-ConvLSTM is also a sequence to sequence network that can predict long time evolution of a dynamical system by passing hidden states from one encoder to another. The network performed well in predicting the dynamic evolution of unsteady 2-D viscous Burgers when trained using data and, in another case, using governing equation (without data), i.e., physics-constrained. Further, AE-ConvLSTM was used in an effort to predict the time evolution of two unsteady Navier-Stokes problems. These problems have coupled pressure and velocity field having different magnitude order, and these fields evolve in time at a different rate. It was observed that the network could be trained using data, but while training using physics-constrained via governing equations, AE-ConvLSTM fails to train for time evolution.

\end{abstract}

\begin {keyword}
Auto-encoder, ConvLSTM, physics-constrained, dynamical systems, Navier-Stokes
\end{keyword}

\end{frontmatter}
\section{Introduction}
\indent Dynamical system modeling has been researched extensively in the past. For most practical problems, dynamical systems governed by nonlinear partial differential equations (PDEs) are difficult to solve analytically. Thus, numerical modeling of these simulations is an ad-hoc approach. Numerical simulations with high accuracy, complex problems, and larger domains require significant computational resources and usually have a big turnaround time which is a bottleneck for industrial applications. After the advent of deep learning, a new approach for predicting dynamical systems has attracted the attention of researchers in recent years~\cite{wang2021physics}. This approach involves replacing or aiding the simulator by creating neural network surrogate models. These models which replaces a numerical simulator should be able to provide solutions for different initial and boundary conditions at a fraction of time to justify the cost of training a network for that particular problem. Once trained, the surrogate model could be an order of magnitude faster in terms of simulation time, which could be helpful in problems like optimization or inverse problems where many simulations are required. When surrogate model is trained using data obtained from a simulator, the training method is called as data-driven method~\cite{brunton2020special}, and when training is done using governing equations, the method is called as physics-constrained method~\cite{raissi2018deep, Geneva2020Modeling, Ren2021Phycrnet}.\\
\indent The philosophy underlying the data-driven model using deep learning is based on the assumption that, given an extensive data set, we can train a nonlinear function, which maps inputs to its corresponding output. This function, i.e., a neural network, can then be used to predict outcomes given corresponding inputs. Based on this assumption, a data-driven surrogate model for a dynamical system can be constructed. Sometimes, the simulation cost for obtaining data is too high, and for creating a surrogate model, a loss function constraint by physics can be used~\cite{raissi2018deep}. Hence, we use governing equations instead of data to calculate the loss, which is minimized during training.\\
\indent As dynamical systems vary in space and time, a neural network should be able to capture spatial-temporal features. In the past, convolutional neural network architectures have been used to capture the spatial characteristics of complex systems~\cite{zhu2018bayesian}. A temporal dimension to these complex systems was introduced by firstly using an auto-regressive network~\cite{Geneva2020Modeling, Mo2019Deep} and later by using recurrent neural networks(RNN) where the time dependencies are maintained through the introduction of additional parameters~\cite{Maulik2020Recurrent, Geneva202Transformers}. A network that combines the convolutional network and RNN is ConvLSTM~\cite{Shi2015CLSTM}. ~\citet{Tang2020} predicted subsurface flows using an encoding decoding net along with ConvLSTM, where a single initialization could predict non-uniform time steps. The authors showed that the network performed better than the auto-regressive network of~\citet{Mo2019Deep}, but the model was limited to ten-time frames. Recently, ~\citet{Ren2021Phycrnet} used ConvLSTM networks to incorporate coupled RNN and convolutions networks for dynamical systems and trained the network using physics-constrained loss. However, the authors predicted a single-time step output in each pass through a neural network given a set of inputs. ~\citet{Mohan2019Compressed} used auto-encoder along with sequence to sequence structure of ConvLSTM for creating a surrogate for homogeneous isotropic turbulence. The authors used data for training 20 snapshots in the future but did not test the model for predicting longer time series.\\
\indent Sequence to sequence model leads to the prediction of multiple time steps in one forward pass through a neural network. Hence, the need to back-propagate through many network graphs is decreased, alleviating the issue of vanishing gradients. This article investigates the ability of a novel sequence to sequence ConvLSTM network, i.e., auto-encoder ConvLSTM (AE-ConvLSTM), to predict the evolution of a dynamical system for a long-time period ~100 steps. The article evaluates the ability of AE-ConvLSTM to be trained on data and on physics-constrained loss (without any data) to predict the evolution of different dynamical systems for long periods especially unsteady Navier-Stokes which is of keen interest to industry~\cite{hennigh2021nvidia}. \\
\indent The structure of the present article is as follows: Section~\ref{sec: AE-ConvLSTM} explains the baseline ConvLSTM model of~\citet{Shi2015CLSTM} and a novel sequence to sequence auto-encoder convolutional LSTM, i.e., AE-ConvLSTM network, is introduced, where its prediction capabilities are compared with the baseline ConvLSTM. Section~\ref{sec: Training Methodology} defines training methodologies used by AE-ConvLSTM network for predicting long periods of evolution in dynamical systems. The two methodologies evaluated here are the data-driven method and the physics-constrained method. Section~\ref{sec: Results} showcases the performance of the data-driven approach and physics-constrained method in predicting the time evolution of the unsteady 2-D viscous Burgers problem and two unsteady Navier-Stokes problems, i.e., lid-driven cavity and a dummy problem of periodic vorticity formulation~\cite{Li2020Fourier}. Final section~\ref{sec: conclusion} concludes this article. The code and the network is available on GitHub~\footnote{\url{https://github.com/kakkapriyesh/AE-ConvLSTM-Flow-Dynamics}}


\section{AE-ConvLSTM}\label{sec: AE-ConvLSTM}
\indent Our goal is to solve physical systems which evolve in space and time. These problems could be governed by ordinary or a partial differential equation in the form
\begin{equation}\label{eqn:Problem Formulation}
\begin{split}
u_{t} = \mathcal{F}\left({x}, \nabla_{x}u, \nabla_{x}^{2}u,c, u_{x}, u_{t}\right), \\
t \in \mathcal{T} \subset \mathbb{R}^{+}, \quad {x} \in \Omega \subset \mathbb{R}^{m},
\end{split}
\end{equation}
where subscripts denote partial derivatives, the domain $\Omega$ consists of a solution $u \in \mathbb{R}^{n\times d}$ in time interval $\mathcal{T}$ and is bounded by a boundary condition $\mathcal{B}$, which could be a function of space and time.  This formulation incorporates a large set of problems from fluids, combustion, and mechanics. \\
\indent Most problems arising from the above formulation are computed numerically where the continuous solution is discretised in both the spatial and the temporal domain i.e., the differential equation for a variable $u$ is discretized as ${u}$=$\left(u_{0}, u_{1}, \ldots u_{T}\right) ; u_{i} \in \mathbb{R}^{n \times d}$ for n variables in interval time $\mathcal{T}$ with a time step of $\Delta{t}$. We can create a surrogate using neural networks instead of a numerical simulator to predict time series for the described physical systems. We propose to create a surrogate model using ConvLSTM layers.
\subsection{Convolutional long-short term memory model (ConvLSTM)}
\begin{figure}
		\centering
		\includegraphics[width=1.0\linewidth]{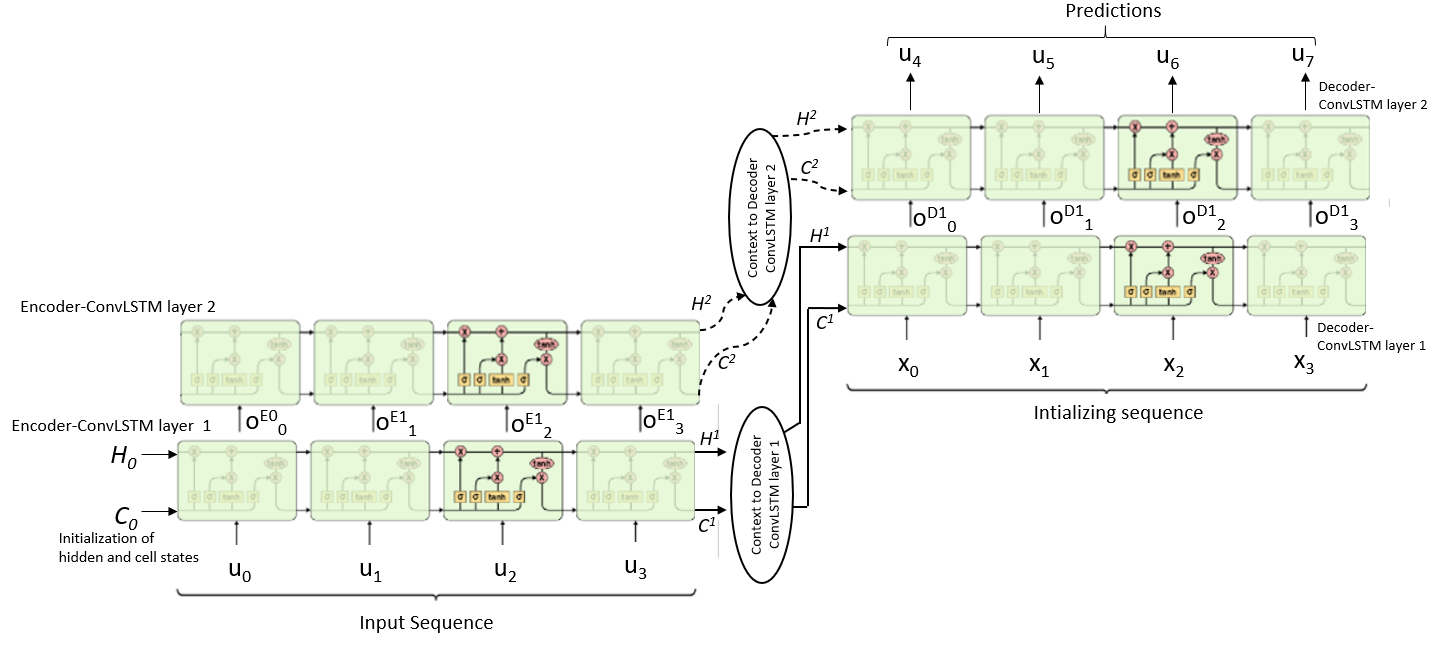}
    	\caption{Encoder-Decoder Structure (Rolled-out)}
    	\label{encoder-Decoder Block graph}
	\end{figure}

\indent ConvLSTM has a recurrent neural network component usually used in natural language processing (NLP), where the sequence of words is correlated by a hidden state. These hidden states encode the inputs from their current state and influence the next state. These simple RNN structures still suffer from the problem of vanishing gradients while training for a very long sequence. Thus, to enable long-term predictions, Long-short term memory(LSTM)~\cite{Hochreiter1997Long} was used. In LSTM, apart from hidden-state $\mathcal{H}$ a cell-state $\mathcal{C}$ is introduced, whose function is to save the relevant long-term information, mitigating the issue of vanishing gradients in the network. Using this principle of LSTM and RNNs, Shi et al.~\cite{Shi2015CLSTM} used a convolutional matrix to map hidden and cell states with inputs $\mathcal{X}$ to create new hidden and cell states. Henceforth, this mapping of hidden and cell states with the sequence of inputs is called a ConvLSTM layer which can be mathematically described by equation~\ref{eqn:ConvLSTM}, where $*$ is the convolutional operator, and $\circ$ is the Hadamard product.
	    

\begin{equation}\label{eqn:ConvLSTM}
\begin{aligned}
i_{t} &=\sigma\left(W_{x i} * \mathcal{X}_{t}+W_{h i} * \mathcal{H}_{t-1}+W_{c i} \circ \mathcal{C}_{t-1}+b_{i}\right), \\
f_{t} &=\sigma\left(W_{x f} * \mathcal{X}_{t}+W_{h f} * \mathcal{H}_{t-1}+W_{c f} \circ \mathcal{C}_{t-1}+b_{f}\right), \\
\mathcal{C}_{t} &=f_{t} \circ \mathcal{C}_{t-1}+i_{t} \circ \tanh \left(W_{x c} * \mathcal{X}_{t}+W_{h c} * \mathcal{H}_{t-1}+b_{c}\right),\\
o_{t} &=\sigma\left(W_{x o} * \mathcal{X}_{t}+W_{h o} * \mathcal{H}_{t-1}+W_{c o} \circ \mathcal{C}_{t}+b_{o}\right), \\
\mathcal{H}_{t} &=o_{t} \circ \tanh \left(\mathcal{C}_{t}\right).
\end{aligned}
\end{equation}
Here $\sigma$ denotes the sigmoid function, $W$ denotes weights, $b$ stands for bias and gates are represented by $f_{t}$, $i_{t}$, $o_{t}$.\\
\indent The introductory ConvLSTM paper~\cite{Shi2015CLSTM} proposed an encoder-decoder model for predicting ten future output time steps based on ten input sequence of images. This sequence to sequence formulation is inspired from machine translation in natural language processing (NLP). The book by~\citet{NLP_stanford} refers to the sequence of words fed into the network one by one in the encoder without any output and creates the hidden representation of all the input sequences as "context". The decoder can use the context in various ways to create a sequence of output. ConvLSTM model uses a similar stacked encoder-decoder network where, n snapshots in a time series $\left(u_{0}, u_{1}, \ldots u_{n}\right)$  is fed as input and m future predictions $\left(u_{n+1}, u_{n+2}, \ldots u_{m}\right)$  as a series of time-steps is obtained as output. A schematic of an encoder-decoder structure is presented in figure~\ref{encoder-Decoder Block graph}, where each green block represents a ConvLSTM layer. 
\subsubsection{Encoder structure in ConvLSTM}
\begin{figure}
	\centering
		\includegraphics[width=0.8\linewidth]{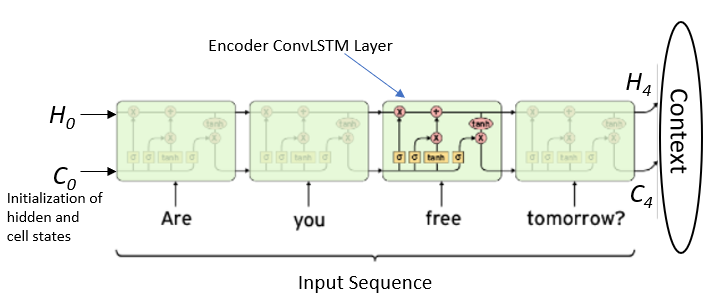}
		\caption{Encoder layer}
		\label{single encoder layer}
\end{figure}

\begin{figure}
	\centering
		\includegraphics[width=0.8\linewidth]{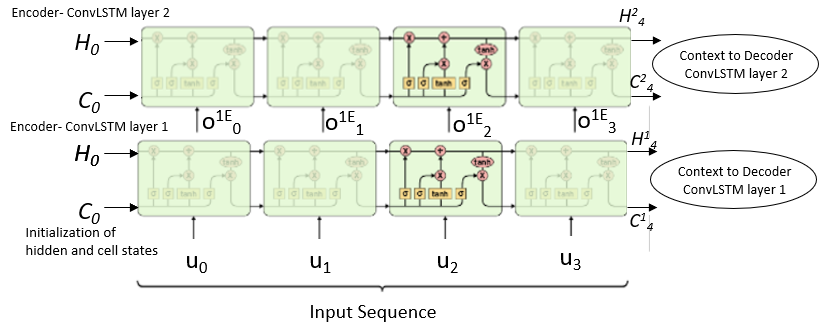}
    	\caption{Encoder structure}
    	\label{Encoder struture}
\end{figure}

\indent For elaborate explanation, a single ConvLSTM encoder layer can be represented as a rolled-out version of the ConvLSTM unit in time, as shown in figure~\ref{single encoder layer}. The time series of input $u_{t}$ is fed to the same ConvLSTM unit sequentially. A hidden state $\mathcal{H}$ and a cell state $\mathcal{C}$ are initialized at the first time step and then are updated one by one with new inputs $\mathcal{X}$  as described by~\citet{Shi2015CLSTM}. After the last input in the sequence, hidden and cell states represent the entire input sequence in the form of context, which is then fed to the corresponding decoder layer. The encoder layer is then stacked as shown in the figure~\ref{Encoder struture} where the output of the first encoder layer is fed to the corresponding time sequence of the second encoder ConvLSTM layer. The initialization and the updating of hidden and cell states are similar to the first encoder layer forming a new context to be passed to the second decoder layer in the decoder structure.
\subsubsection{Decoder structure in ConvLSTM}
\begin{figure}
	\centering
		\includegraphics[width=0.8\linewidth]{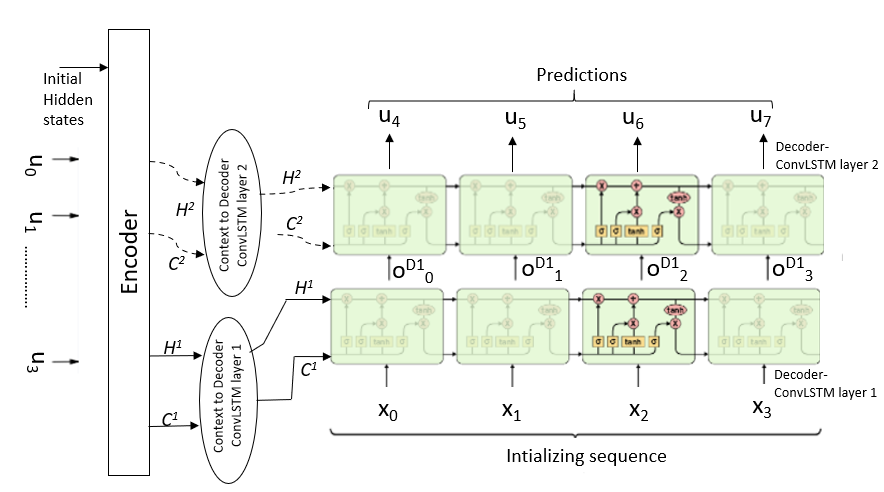}
    	\caption{Decoder structure}
    	\label{Decoder struture}
\end{figure}

\indent Once the context is passed from a ConvLSTM encoder layer, it is fed to the decoder layer as shown in the figure~\ref{Decoder struture}. The hidden and the cell states are then passed through the ConvLSTM layer, where they interact with the inputs of the first decoder layer comprising randomly generated sequence $\mathcal{X}^{r}$. This method of randomly generating input sequence for all the time steps in the decoder is called as an unconditional decoder, which is different from a conditional decoder used in classical machine translation NLP problems~\cite{NLP_stanford}. Classical machine translation problem predominantly uses a sequence to sequence encoder-decoder structure where the context from the encoder is fed to the first step of the decoder ConvLSTM layer, which then uses a randomly generated input to obtain an output. Later, the output from the randomly generated input in the first step of the decoder sequence is fed back as input to the second step of the sequence. The feedback is continued until the sequence is completed. We deviate from this method because it was found that for the prediction of a series of images, there is not a considerable difference between the conditional and unconditional decoder setup when studied by~\citet{Srivastava2015Unsupervised}.
Moreover, the current decoder architecture enables all the output sequences of the first decoder layer to be used as inputs to the next corresponding decoder layer. This makes the architecture at the decoder similar to the encoder leading to a more straightforward implementation. The output of the final decoder layer is the sequence of expected predictions.  

\subsection{Auto-encoder convolutional long-short term memory (AE-ConvLSTM)}\label{Sec: AE-ConvLSTM Full RNN}
\begin{figure}
		\centering
		\subfloat[Schematic for auto-encoder ConvLSTM network]{\includegraphics[width=0.7\linewidth]{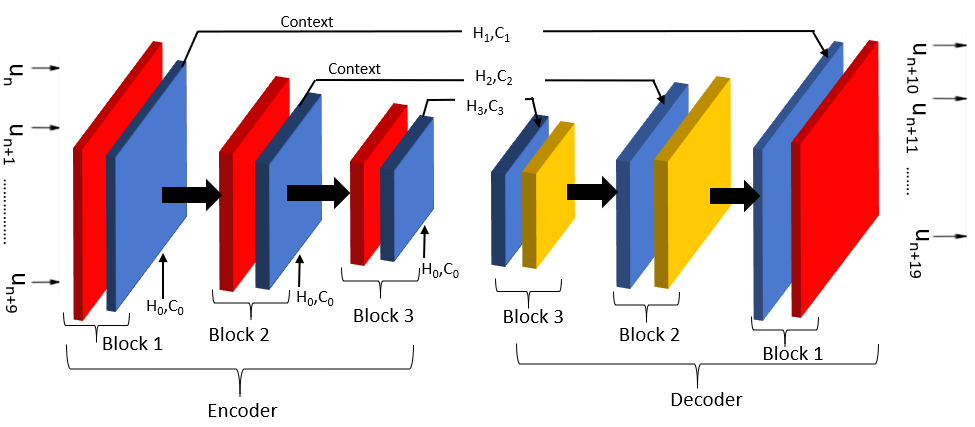}\label{Auto-Encoder Full schematic}}
    	\caption{Red represents classical convolutional layer, blue represents ConvLSTM layer and yellow represents de-convolutional layer.}
\end{figure}

\indent The ConvLSTM model has an encoder-decoder network, but the image's dimensions remain the same throughout the network. We use an auto-encoder network using a classical convolutional layer and the ConvLSTM layer to reduce the dimensions building on the network of the git repository~\cite{gitonline}. The dimensions are reduced with each passing encoder layer, creating a bottleneck; later, the decoder layer increases the dimensions to the original size. This method enables us to use more convolutional filters in the network, making the network deeper without an excessive increase in the parameters. The schematic of the auto-encoder ConvLSTM (AE-ConvLSTM) model is shown in figure~\ref{Auto-Encoder Full schematic}. The AE-ConvLSTM has an encoder of three blocks, where each block comprises a convolutional layer and a ConvLSTM layer. In each encoder block, the image sequence is passed as a stack through every convolutional layer, whereas in the ConvLSTM layer, the images are passed in a series one at a time. The hidden state $\mathcal{H}_{0}$ and cell states $\mathcal{C}_{0}$ are initialized for the ConvLSTM layer in each encoder block, and once the context is formed, those states are passed to the corresponding ConvLSTM layer of the decoder block. The output of one encoder block is then passed as input to another, as shown with a bold black arrow in figure~\ref{Auto-Encoder Full schematic}.\\
\indent Similar to the encoder structure, the decoder structure of AE-ConvLSTM again consists of three blocks. The input to the first block of the decoder is the sequence of random initialization, while the output of the same first block is the sequence fed as input to the next block of the decoder. Each block of the decoder comprises a ConvLSTM layer and a de-convolutional layer which is the transpose of the convolutional layer used for increasing image dimension. The ConvLSTM layer in each block has context as an input from the corresponding encoder block, and the output of the final decoder block is the sequence of predictions.

 \begin{figure}[H]
		\centering
		\subfloat[Baseline ConvLSTM results]{\includegraphics[width=0.8\linewidth]{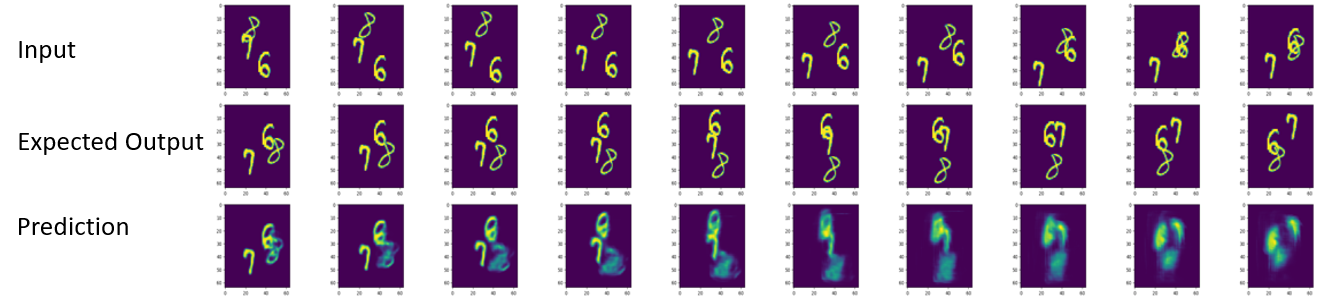}\label{Replication_MNIST1}}
		\vspace{3mm}
	\subfloat[Results from auto-encoder
	ConvLSTM ]{\includegraphics[width=0.8\linewidth]{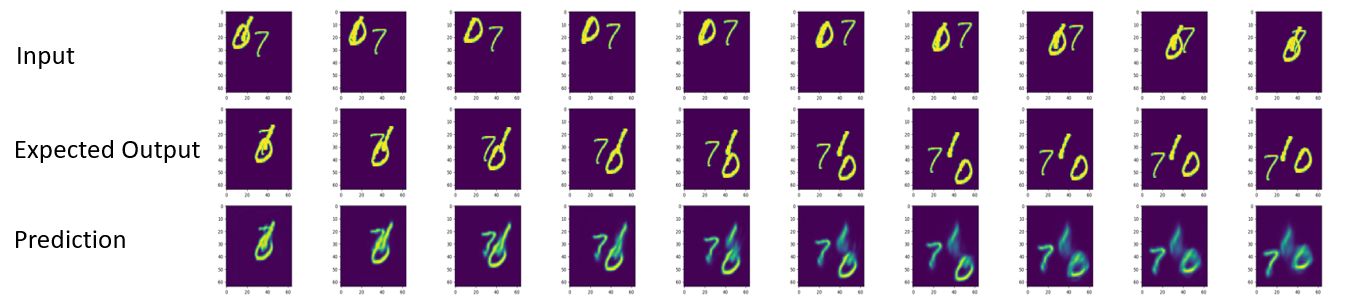}\label{ED_MNIST}}\\
 
	\caption{}
\end{figure}

\begin{table}[H]
    \caption{‘$5\times5$’ and ‘-$3\times3$’ represent the corresponding state-to-state kernel size. $5\times5$ are LSTM layers while $3\time3$ are the vanilla CNN layers. ‘128’, ’64’ , ‘96’ and ‘16’ represents the outer channel in all the layers and () indicates square image size in both height and width.}
    
    \begin{tabular}{l|llcc}
    \hline
               & \multicolumn{1}{c}{Model} & \Longstack{ Parameters} &  \\ \hline
    \Longstack{Baseline ConvLSTM}   & \Longstack{\\(64)-$5\times5$-128(64)-$5\times5$-64(64)\\-$5\times5$-64(64)}          & $ 7.8 \mathrm{mn}$        \\ \\
    AE-ConvLSTM & \Longstack{ (64)-$3\times3$-16(64)-$5\times5$-64(64)-$3\times3$-\\ 64(32)-$5\times5$-96(32)-$3\times3$-96(16)-$5\times5$-96 }           & $ 9.0 \mathrm{mn}$ \\  \hline  \end{tabular}
    \label{tab:Comparison MNIST}
\end{table}
\indent The usefulness of the classical auto-encoders in allowing the network to ignore the noise and enabling the use of deeper layers due to dimensional reduction could be is seen in figure~\ref{ED_MNIST}. Here, the moving MNIST dataset has ten frames as input, and ten frames in a sequence are predicted as output. It can be seen that the performance of AE-ConvLSTM improves significantly as compared to the baseline ConvLSTM model for sequence to sequence predictions. The model parameters used for the baseline ConvLSTM similar to ~\citet{Shi2015CLSTM} and the present AE-ConvLSTM models are shown in table~\ref{tab:Comparison MNIST}.
\subsubsection{AE-ConvLSTM for long term predictions}
\begin{figure}[H]
		\centering
		\includegraphics[width=\linewidth]{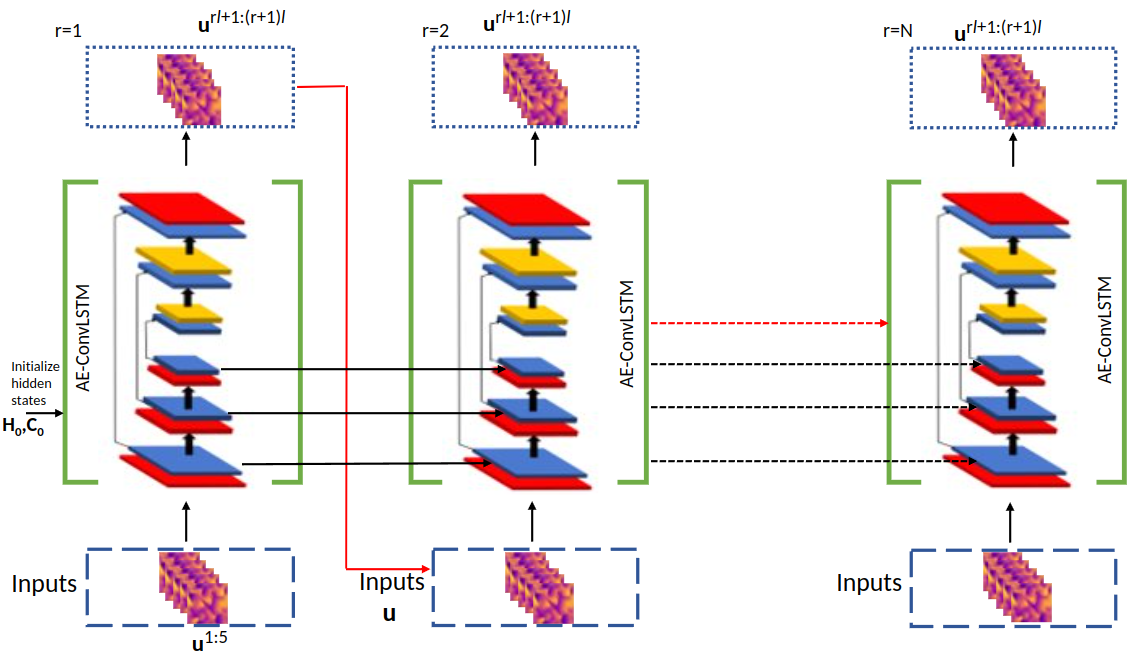}
		\vspace{3mm}
	\caption{Hidden states being passed from one AE-ConvLSTM Module's encoder to another (Module is rolled out). r is the Module sequence number, and $l$ is number of prediction in a single Module.}
	\label{Full RNN}
	\end{figure}

\indent The previous sections discussed the AE-ConvLSTM and the baseline ConvLSTM models for the sequence to sequence prediction. We call one such sequence to sequence model a ``module''. These modules can be extended to learn long-term predictions, i.e., a single sequence can be used as an input to obtain multiple sequences as output. The hidden states and cell states passed from the encoder block to the decoder block within a module help predict a sequence of predictions. However, for long-term predictions, the hidden and cell states can be passed from an encoder block of one module to the encoder block of the next module. The schematic for these long-term predictions with ``module'' rolled out is shown in figure~\ref{Full RNN}.\\
\indent Algorithm for predictions through AE-ConvLSTM is outlined in Algorithm~\ref{alg: pred_algo_dd}.  
\begin{algorithm}
    \caption{AE-ConvLSTM predictions }

    \label{alg: pred_algo_dd}
    \KwIn{$\bf{Input}$: Trained neural-network model: $\mathcal{N}(\cdot, {w})$; sequence length: $\textit{l}$, initial test data: ${\overline{u}_{1:\textit{l}}}$; max number of time-steps to predict: $T_{max}$, hidden state $\mathcal{H}_{0}$ and cell state $\mathcal{C}_{0}$.}
    ${\chi}_{1} \leftarrow \left\{{\overline{u}_{1}}, {\overline{u}_{2}}, \ldots, {\overline{u}_{\textit{l}}},\mathcal{C}_{0},\mathcal{H}_{0}\right\}$;
   
    \For{$r$ = $1$ \KwTo $N_{seq}$= {$T_{max}$}/{$\textit{l}$}}
    {
        ${u}_{(r\textit{l}+1):(r+1)\textit{l}} \leftarrow \mathcal{N}\left({\chi}_{r},\text{w}\right)$; \Comment{Forward pass of model}\\
        ${u}_{out}[r] \leftarrow {u}_{{(r\textit{l}+1):(r+1)\textit{l}}};$\\
        ${\chi}_{r+1} \leftarrow \left\{{u}_{{(r\textit{l}+1):(r+1)\textit{l}}},
        {\chi}_{r}\right\}$\\\Comment{Update input, cell states and hidden states}
    }
\KwOut{\text { Predicted time series } ${u_{\text {out }}}=\left[ {u_{\textit{l}+1}}, {u_{\textit{l}+2}}, \ldots, {u_{T_{\mathrm{max}}}}\right]$}
\end{algorithm}
The formulation of AE-ConvLSTM can be written as a network built using successive layers of convolution layers and the ConvLSTM layers in the encoder and decoder which can be parameterized as the function composition: 
\begin{equation}
  \begin{array}{l}
\mathcal{N}\left(\chi^{d}, {w},\mathcal{C}^{d},\mathcal{H}^{d}\right)=\operatorname{decoder} \circ \operatorname{encoder}\left(\chi^{d},\mathcal{C}^{d},\mathcal{H}^{d}\right), \\
{u}^{d}=\operatorname{decoder}\left({z}, {w}_{d},\mathcal{C}^{d},\mathcal{H}^{d}\right), \quad{z},\mathcal{C}^{d},\mathcal{H}^{d}=\operatorname{encoder}\left(\chi^{d}, {w}_{e},\mathcal{C}^{0},\mathcal{H}^{0}\right)
\end{array},  
\end{equation}
where $\left\{{w}_{e}, {w}_{d}\right\}={w}$ are the encoder and decoder parameters, respectively. ${z}$ are latent variables that are of lower dimensionality than ${u}^{d}$. The input channel dimensionality of the convolutional model is the product of the number of state variables $d_{0}$ and the sequence length $\textit{l}$. The number of output channels is equivalent to the number of state variables predicted and the sequence length. While we use $\mathcal{N}\left(\chi^{d}, {w},,\mathcal{C}^{d},\mathcal{H}^{d}\right)$ to encapsulate this process, one can interpret this model as learning two processes: encoding data from previous time-steps to a latent variable $z$ and the prediction of the next state as a function of $z$ with hidden and cell states remembering long term representation.\\
\section{Training methodology}~\label{sec: Training Methodology}
\subsection{Data-driven method (Training with data)}
\label{subsec: Train data}
\begin{figure}[H]
		\centering
		\includegraphics[width=0.9\linewidth]{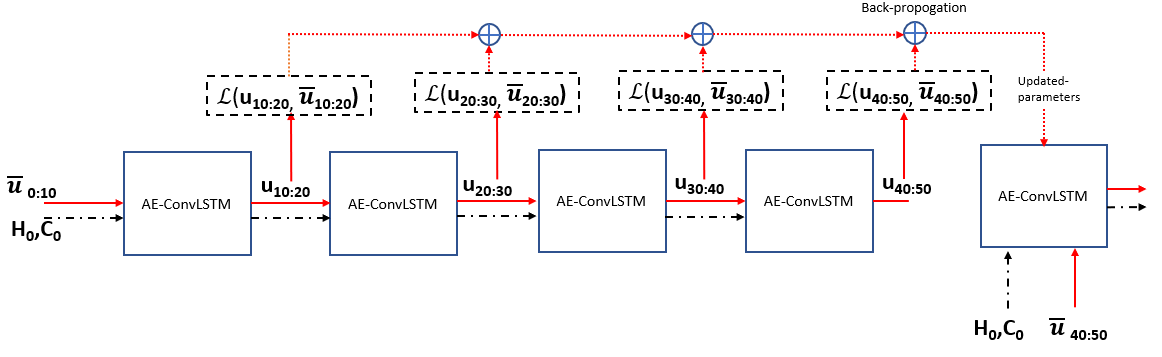}
	\caption{Multi time-step back-propagation of AE-ConvLSTM where ${u_{n}}$ is the output prediction, ${\overline{u}_{n}}$ is the data obtained from the simulator and sequence length $\textit{l}=10$. }
	\label{Train_FF}
	\end{figure}

\indent The data-driven training method requires data to be generated in the simulator with varying initial conditions for a predefined period of simulation time. These generated data are used to train the neural network. For AE-ConvLSTM, the input is a stack of field snapshots with a predefined sequence length ${l}$ as shown in the figure~\ref{Train_FF}. The data or feature maps ${u}_{(r\textit{l}+1:(r+1)\textit{l})}$ are passed through convolutional layers as described in section~\ref{Sec: AE-ConvLSTM Full RNN}, and after each pass through neural-network, the output of the neural-network ${u}_{(r\textit{l}+1:(r+1)\textit{l})}$ is evaluated against the corresponding data $\overline{{u}}_{(r\textit{l}+1:(r+1)\textit{l})}$. The network is then trained with a gradient descent optimization algorithm to minimize the mean-squared error loss posed as follows,
\begin{equation}\label{eq: loss_dd}
    \mathcal{L}{\left({u},\overline{{u}}\right)}=\frac{1}{M} \sum_{j=1}^{M} \sum_{i=1}^{p \times l}\left\|\overline{{u}}_{i}^{j}-{u}_{i}^{j}\right\|_{2}^{2},
\end{equation}
here M is the mini-batch and $p \times l$ is the number of time-steps predicted before back-propagation occurs. 
The hidden state $\mathcal{H}$ and cell state  $\mathcal{C}$ are initialized once per back-propagation as shown in figure~\ref{Train_FF} and are fed forward to the next training loop. After summing the losses for predefined modules {p}, we back-propagate through all the modules to update weights. The network with updated weights is again trained forward in time with new initialized hidden and cell states along with the data from the simulator for the corresponding time step. The sequence length $\textit{l}$ and modules stored before back-propagation is a hyper-parameter p which can be tuned to obtain better results. The training process is outlined in Algorithm~\ref{alg: train_dd}.\\
\begin{algorithm}
    \caption{Training AE-ConvLSTM (data-driven) }
    \label{alg: train_dd}
    \KwIn{ Neural-network model: $\mathcal{N}(\cdot, {w})$; Sequence length: $\textit{l}$, Training data: ${\overline{u}}$; number of time-steps to unroll: $T_{max}$., Back-prop interval: $p$, Learning rate: $\eta$, Hidden state $\mathcal{H}_{0}$ and cell state $\mathcal{C}_{0}$, Number of epochs: ${N}$.}
     \For{$\textrm{epoch} = 1$ \KwTo $N$}{
    ${\chi}_{1} \leftarrow \left\{{\overline{u}_{1}}, {\overline{u}_{2}}, \ldots, {\overline{u}_{\textit{l}}},\mathcal{H}_{0},\mathcal{C}_{0}\right\}$\\
    \For{$r$ = $1$ \KwTo $N_{seq}$= {$T_{max}$}/{$\textit{l}$}}
    {
        ${u}_{(r\textit{l}+1):(r+1)\textit{l}} \leftarrow \mathcal{N}\left({\chi}_{r},{w}\right)$; \Comment{Forward pass of model}\\
        \vspace{2 mm}
         $\mathcal{L}_{r}$ = $\mathcal{L}_{r-1}$ + $\textrm{MSE}\left(\overline{{u}}_{(r\textit{l}+1):(r+1)\textit{l}}, {u}_{(r\textit{l}+1):(r+1)\textit{l}}\right)$\\ \Comment{Calculate Loss}\\
         \If{Mod(r,p)=0}{ 
                $\nabla {w} \leftarrow  \textrm{Backprop}(\mathcal{L}_{r})$ \Comment{Multi-step back-propagation}
                ${w} \leftarrow {w} - \eta \nabla {w}$ \Comment{Gradient Descent}\\
                 $\mathcal{L}_{r} = 0$ \Comment{re-intialize loss}
            }
        ${\chi}_{r+1} \leftarrow \left\{{u}_{(r\textit{l}+1):(r+1)\textit{l}},
        {\chi}_{r}\right\}$\\ \Comment{Update input}
    }}
    \KwOut{Trained AE-ConvLSTM model $\mathcal{N}\left(\cdot,{w}\right)$;}
\end{algorithm}

\subsection{ Physics-constrained method (Training using governing equation)}
\label{subsec: Train physics Burgers}
\begin{figure}[H]
		\centering
		\includegraphics[width=0.9\linewidth]{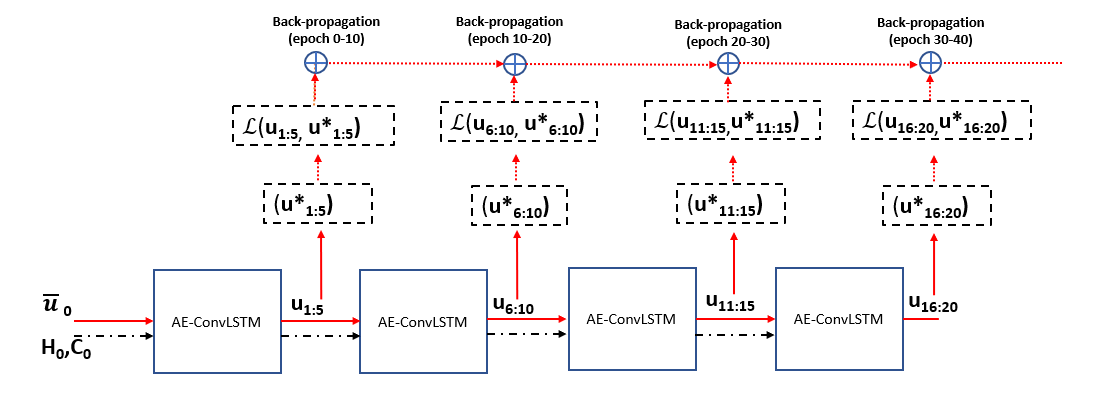}
		%
	\caption{Multi time-step back-propagation of AE-ConvLSTM where ${u_{n}}$ is the output prediction from the model, ${{u^{*}_{n}}}$ is the next time-step obtained from governing equations, and sequence length $\textit{l}=5.$}
	\label{Train PC}
	\end{figure}

\indent Training the neural network using physics-constrained method involves using the system's governing equations as constraints during the optimization step (i.e., training is constrained by physics instead of data). The initial field of the dynamics is fed as input, and the neural network is trained for the system's future evolution in time by optimizing loss function using the governing PDEs. These governing PDEs are introduced into the loss function by discretizing it in space and time using the finite difference method. First, we discretize the equations in time using implicit methods like Crank-Nicholson, which leads to the evolution of time from $t$ to $t+\Delta{t}$. The domain $\Omega$ is discretized using the second-order accurate central difference scheme. The evolution of field ${u}$ can be described as a time-evolving PDE given by
\begin{equation}
\label{eq:CN}
  \frac{{{u}^{*}}_{n+1}-{u}_{n}}{\Delta t}=\frac{1}{2}\left[F_{\Delta x}\left({x}, {u}_{n}\right)+F_{\Delta x}\left({x}, {u}_{n+1}\right)\right], 
\end{equation}
while the unknown field $ {u}_{n+1}$ in equation~\ref{eq:CN}, is obtained from the neural network when previous fields $\chi_{n}$ is given as input
\begin{equation}
    {u}_{n+1}=\mathcal{N}\left(\chi_{n}, {w}\right), \quad \chi_{n} \equiv\left\{{u}_{n}, {u}_{n-1}, \ldots, {u}_{n-k}\right\}.
\end{equation}
 ${u}^{*}_{n+1}$ in equation~\ref{eq:CN} is the output of the governing equation
 and thus the loss function is difference between the future time step predicted by the neural network and the governing PDE:
 \begin{equation}
     \mathcal{L}=\frac{1}{M} \sum_{i=1}^{M}\left\|{u}^{i}_{n+1}-{{u}^*}^{i}_{n+1}\right\|_{2}^{2}.
 \end{equation}
\indent This loss is minimized to ensure that the neural network is trained to produce the output that satisfies systems governing equations. Physical constraints for fluid systems have been done in the past by~\cite{Geneva2020Modeling, Ren2021Phycrnet} but we use sequence to sequence prediction instead of predicting a one-time step per forward pass of the neural network. The training process for the AE-ConvLSTM network is described in figure~\ref{Train PC}. The network is given an initial field ${\overline{u}}$ as an input along with the hidden state $\mathcal{H}_{0}$  and the cell states $\mathcal{C}_{0}$. The network then predicts next sequence of time steps ${u}_{{(r\textit{l}+1):(r+1)\textit{l}}}$. Each output is then fed to the governing equations of the PDE as described earlier to minimize the loss. The training of this network using physics-constrained method is difficult hence, the network rollout is incremented after certain epochs $p$, which is a hyperparameter. This slow rollout enables the network to learn the dynamics slowly, as explained in the pseudocode described in algorithm~\ref{alg: train_PC}. The number of epochs $N$ and the number of time steps to be trained $T_{max}$ are predetermined, and the number of networks to be rolled out after which back-propagation occurs, i.e., back-prop interval $p$ is increased gradually as epochs increase. Once the back-propagation is completed, the weights are updated, and the training is reset with a different initial input field. 

\begin{algorithm}[H]
    \caption{Training AE-ConvLSTM (physics-constrained method) }
    \label{alg: train_PC}
    \KwIn{Neural-network model: $\mathcal{N}(\cdot, {w})$; sequence length: $\textit{l}$, initial field: ${\overline{u}_{0}}$; max number of time-steps to unroll: $T_{max}$., learning rate: $\eta$, hidden state $\mathcal{H}_{0}$ and cell state $\mathcal{C}_{0}$, discretized RHS $F_{\Delta x}$ for spatial derivatives, number of epochs: ${N}$, $\textrm{tsteps} = \textrm{linspace}\left(1, T_{max}, N\right)$.}
     \For{$\textrm{epoch} = 1$ \KwTo $N$}{
    ${\chi}_{1} \leftarrow \left\{{\overline{u}_{0}}, {\overline{u}_{0}}, \ldots, {\overline{u}_{\textit{0}}},\mathcal{H}_{0},\mathcal{C}_{0}\right\}$
    
    $p \leftarrow \textrm{tsteps}[\textrm{epoch}]$ \tcp{Time-steps to unroll}
    \For{$r$ = $0$ \KwTo $T$}
    {
        ${u}_{(r\textit{l}+1):(r\textit{l}+\textit{l})} \leftarrow \mathcal{N}\left({\chi}_{r},{w}\right)$; \tcp{Forward pass of model}
        \vspace{2 mm}
        \For{$i$ = $(r\textit{l}+1)$ \KwTo $(r\textit{l}+\textit{l})$}{
          ${{u}}^{*}_{i} = T_{\Delta t}\left({u}_{i}, F_{\Delta x}\right)$ \hspace{1cm}\tcp{Time integration}
            $\mathcal{L}_{i}$ = $\mathcal{L}_{i-1}$ + $\textrm{MSE}\left({{u}}^{*}_{i}, {u}_{i}\right)$ \hspace{1cm}\tcp{Calculate loss}}
         
          ${\chi}_{r+1} \leftarrow \left\{{u}_{(r\textit{l}+1):(r\textit{l}+\textit{l})},
        {\chi}_{r}[0], {\chi}_{r}[1], \ldots, {\chi}_{r}[k-1]\right\}$ \hspace{1cm}\tcp{Update input}}
                $\nabla {w} \leftarrow  \textrm{Backprop}(\mathcal{L}_{i})$ \hspace{1cm}\tcp{Multi-step back-prop}
                ${w} \leftarrow {w} - \eta \nabla {w}$ \hspace{1cm}\tcp{Gradient descent}
                 $\mathcal{L}_{i} = 0$ 
            
       
    }
    \KwOut{Trained AE-ConvLSTM model $\mathcal{N}\left(\cdot,{w}\right)$;}
\end{algorithm}
\noindent For all the physics-constrained cases attempted in this article, no data accept the initial field were used for training the network. 
\section{Results}~\label{sec: Results}
\indent AE-ConvLSTM network discussed in the above section is tested on various flow problems for both the data-driven and the physics-constrained cases. This section discusses the governing equations of the problems attempted and compares the model's performance between the data-driven case, the physics-constrained case, and previous literature.
\subsection{2-D coupled viscous Burgers equation}
\subsubsection{Data-driven method}\label{sec train_Burgers}
\begin{figure}[H]
    \centering
    \includegraphics[width=\textwidth]{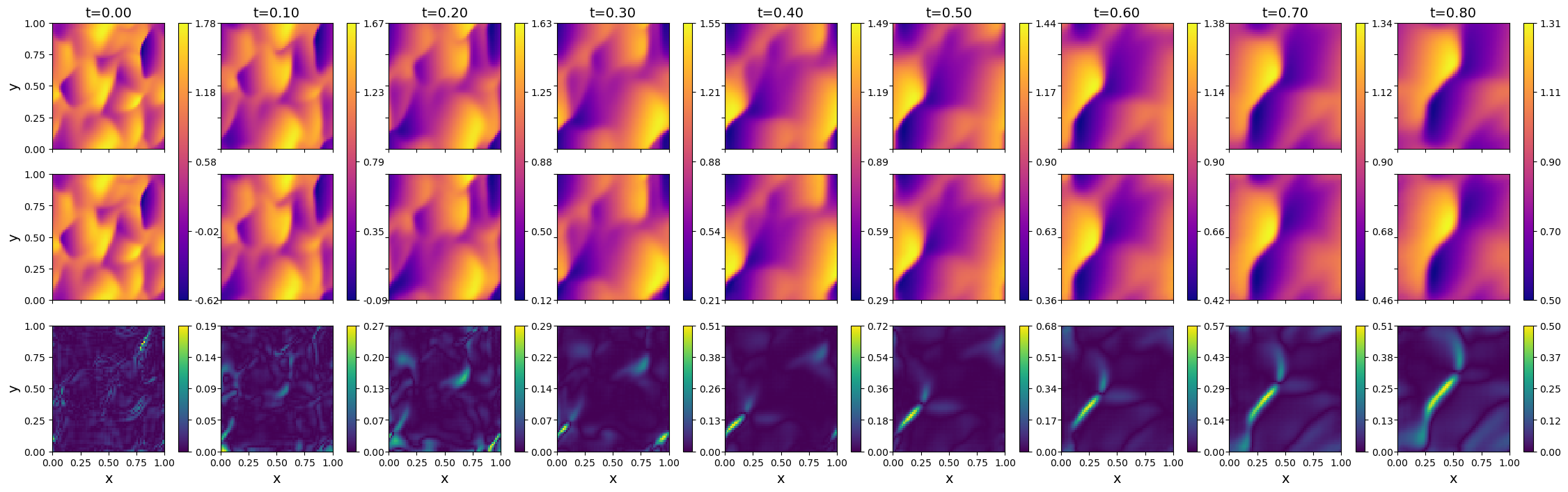}
    \caption{AE-ConvLSTM predictions of a 2D coupled Burgers' test case. (Top to bottom) $x$-velocity FEM target solution, $x$-velocity AE-ConvLSTM prediction, $x$-velocity $L_1$ error.}
    \label{DD_Burgers}
    \end{figure}

\indent The formulation of the coupled Burgers equation is taken from the work of~\citet{Geneva2020Modeling}. The training data used was obtained from a FEM Fenics solver. Each simulation progresses for 100-time steps, with 3600 simulations used as data. The length sequence $l$ used for training is ten, i.e., we predict ten sequences of output with ten inputs. The back-propagation interval $p$ described in section~\ref{subsec: Train data} is kept as 5. i.e., after predicting 50-time steps in training and accumulating loss, the neural network will back-propagate to update weights by calculating gradients. The optimizer used is Adam's, with an initial learning rate of $0.0001$. The learning rate decreases by half if the valid loss does not decrease for two consecutive epochs and the training is done for 100 epochs. The validation is done with 200 simulations. $3.71 \times 10^{6}$ parameters were used in the network with configuration shown in table~\ref{tab:Bugers param}. Figure~\ref{DD_Burgers} showcase's the accuracy of the test prediction done. The model was able to accurately predict the flow features and shock discontinuities across the domain for the whole period of 100-time steps.
\begin{table}[H]
    \caption{‘$5\times5$’ and ‘-$3\times3$’ represent the corresponding state-to-state kernel size. $5\times5$ are LSTM layers while $3\times3$ are the vanilla CNN layers. ‘128’, ’64’ , ‘96’ and ‘16’ represents the outer channel in all the layers and () indicates square image size in both height and width.}
    \centering
    \begin{tabular}{|c|c|c|}
    \hline
       & {Model} &  {Parameters}  \\
    \hline
    {AE-ConvLSTM } & {(64)-$3\times3$-16(64)-$3\times3$-64(64)-$3\times3$-} & \multirow{2}*{$ 3.71 \mathrm{mn}$} \\
    (Burgers) & {$3\times3$-96(32)-$3\times3$-96(16)-$3\times3$-96 } &           \\  \hline  \end{tabular}
    \label{tab:Bugers param}
\end{table}
\subsubsection{Physics-constrained method}
\begin{figure}[H]
    \centering
    \includegraphics[width=\textwidth]{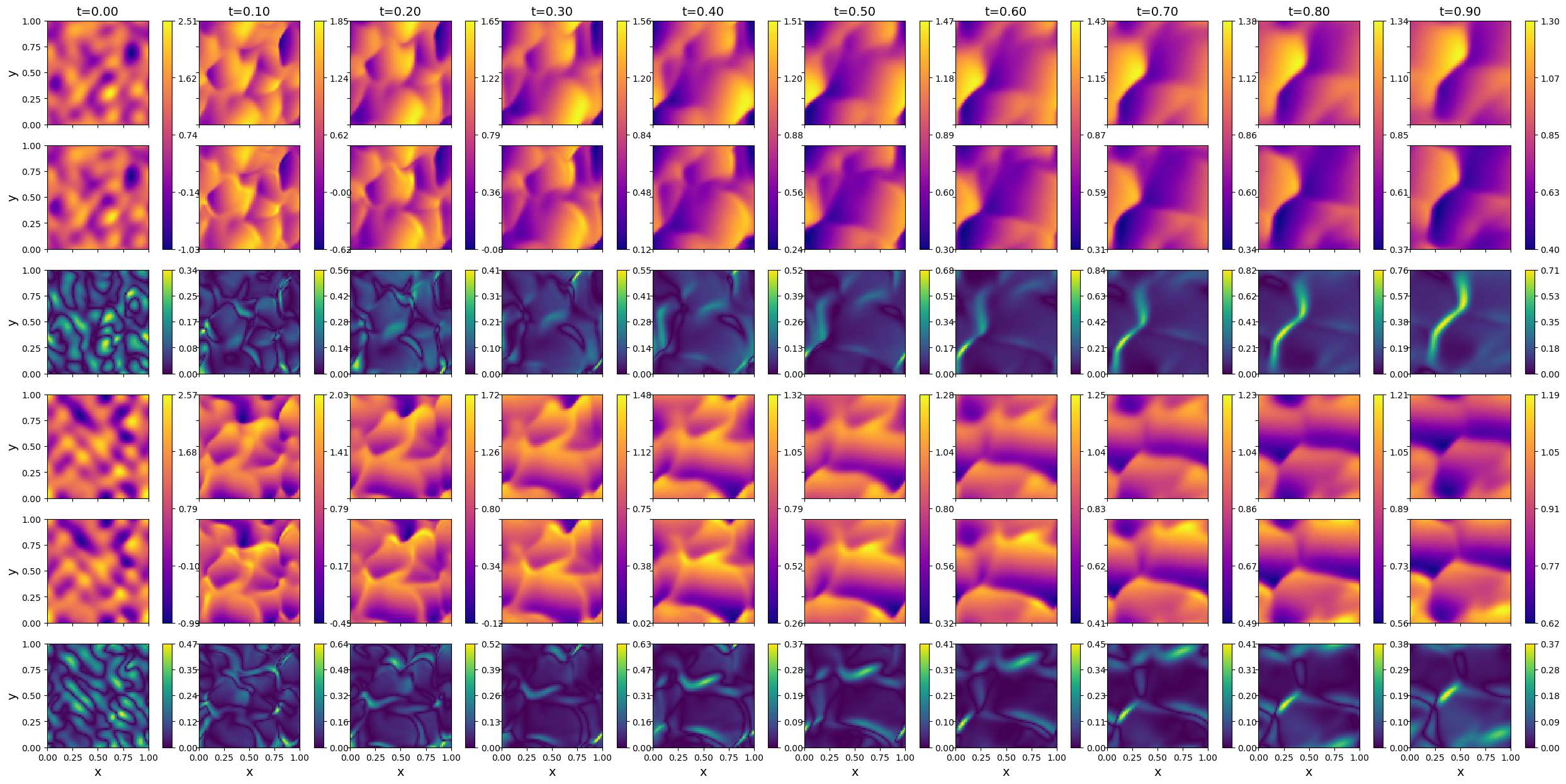}
    \caption{AE-ConvLSTM predictions of a 2D coupled Burgers' test case. (Top to bottom) $x$-velocity FEM target solution, $x$-velocity AE-ConvLSTM prediction, $x$-velocity $L_1$ error, $y$-velocity FEM target solution, $y$-velocity AE-ConvLSTM prediction and $y$-velocity $L_1$ error.}
    \label{PC_Burgers}
\end{figure}

\indent Training using governing equations i.e., physics-constrained is shown in algorithm~\ref{alg: train_PC}. The time step interval for the simulation, i.e., $\Delta{t}$, is taken to be $0.005$ seconds. The sequence length $l$ is five, and the back-propagation interval was gradually increased after every ten epochs, which is described in figure~\ref{Train PC}. The Crank-Nicholson scheme is used for time discretization, and central difference scheme is used for the space discretization. Apart from the change in the choice of the back-propagation interval $p$, all the parameters and hyper-parameters used in the current training methodology are similar to the training done using data. The training was done till the time-step of 0.35 seconds, and the results are extrapolated while testing, as shown in figure~\ref{PC_Burgers}.

\begin{figure}[H]
    \centering
    \includegraphics[width=0.8\textwidth]{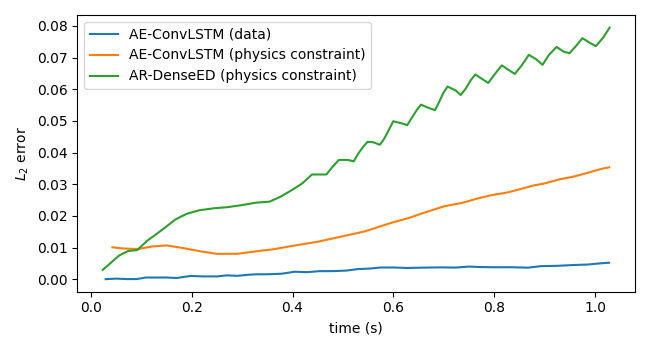}
    \caption{$L_{2}$ error in predictions of a 2D coupled Burgers test case when AE-ConvLSTM is trained with data and physics-constrained method compared against AR-DenseED model~(\citet{Geneva2020Modeling}).}
    \label{mseloss}
\end{figure}

\indent A comparison plot is shown in figure~\ref{mseloss}, where $L_{2}$ error in the predictions done by AE-ConvLSTM network when trained with data and physics-constrained is compared against the predictions done by AR-DenseED model of \citet{Geneva2020Modeling}. AR-DenseED is an auto-regressive model which has encoder-decoder features and uses dense layers~\cite{huang2017densely}. When trained with data, AE-ConvLSTM predicts more accurate solutions than it is trained with physics-constrained method. AE-ConvLSTM performs better probably because it is an RNN network when compared to AR-DenseED, which tries to predict the long-term solutions in an auto-regressive way. AE-ConvLSTM is trained for 0.35 seconds, and the rest is extrapolations, due to which error increases rapidly after 0.35-sec. 

\subsection{Training of neural-network for unsteady Navier-Stokes solution.}
\indent AE-ConvLSTM was able to predict the unsteady 2-D viscous Burgers problem. However, this problem is more straightforward than the unsteady Navier-stokes (N-S) solution, where pressure and velocity fields have different magnitudes. Unlike x and y velocities in the Burgers problem, these coupled pressure and velocity fields evolve at very different rates in time. A couple of test cases i.e., lid-driven cavity and a periodic vorticity formulation by ~\citet{Li2020Fourier} are tried to test the ability of AE-ConvLSTM to predict the unsteady Navier-stokes problem.   
\subsubsection{Lid-driven cavity}

\begin{figure}[H]
		\centering
		\includegraphics[width=1\linewidth]{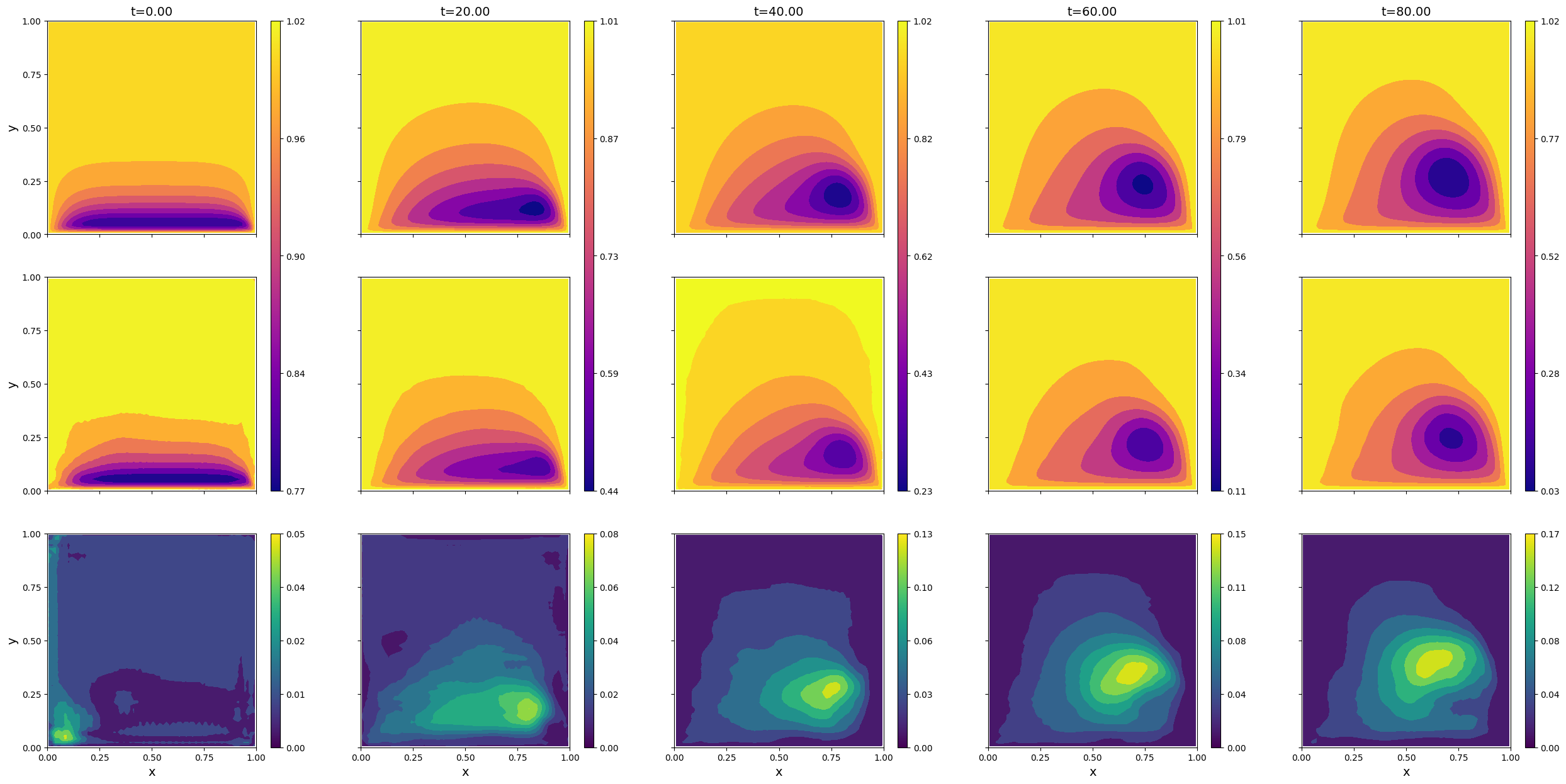}
	\caption{Prediction of the evolution of $\psi$ field in time when trained using data-driven method. (Top to bottom) finite difference target solution, AE-ConvLSTM prediction, $L_1$ error.}
	\label{lid_data_results}
	\end{figure}

\begin{figure}[H]
		\centering
		\includegraphics[width=0.9\linewidth]{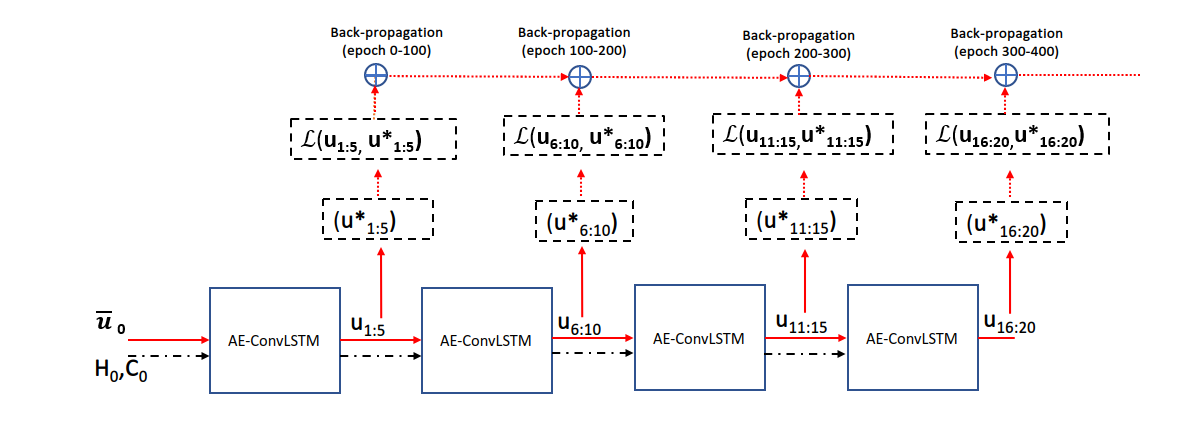}
	\caption{Multi time-step back-propagation of AE-ConvLSTM where ${u_{n}}$ is the output prediction from the model, ${{u^{*}_{n}}}$ is the next time-step obtained from governing equations. }
	\label{Train lid PC}
	\end{figure}

\indent A lid-driven cavity is a canonical problem in fluid mechanics used to validate numerical simulators. The problem involves a 2-dimensional cavity with a lid on top moving at a constant speed, which constantly stirs the fluid at rest. The cavity is discretized uniformly in $x$ and $y$ direction with a $64 \times 64$ grid. The governing equation used in the problem has a stream-vorticity formulation derived from the Navier-Stokes equation given by

\begin {equation}
\begin{array}{l}
\label{eq:lid_laplace}
\frac{\partial^{2} \psi}{\partial x^{2}}+\frac{\partial^{2} \psi}{\partial y^{2}}=-\omega,\\
\end{array}
 \end{equation}
  \begin {equation}
\begin{array}{l}
\label{eqn: vort lid}
\frac{\partial \omega}{\partial t}+\frac{\partial \psi}{\partial y} \frac{\partial \omega}{\partial x}-\frac{\partial \psi}{\partial x} \frac{\partial \omega}{\partial y}=\nu\left(\frac{\partial^{2} \omega}{\partial x^{2}}+\frac{\partial^{2} \omega}{\partial y^{2}}\right).
\end{array}
 \end{equation}
The spatial derivatives are discretized using the central difference scheme while the time derivative is discretized using a Crank-Nicholson scheme. The boundary conditions are derived using equation~\ref{eq:lid_laplace} and Taylor's series expansion. Stream function $\psi$ is considered constant along the walls. The boundary conditions after discretization are given by,
\begin{equation}
 \begin{array}{l}
\text { Top : } \omega_{i, j=n}=2 \frac{-\psi_{i, j-1}}{h^{2}}-\frac{2 U_{top}}{h}, \quad \text { Left : } \omega_{i=1, j}=2 \frac{-\psi_{i+1, j}}{h^{2}}, \\
\text { Bottom : } \omega_{i, j=1}=2 \frac{-\psi_{i, j+1}}{h^{2}}, \quad \text { Right : } \omega_{i=n, j}=2 \frac{-\psi_{i-1, j}}{h^{2}}.
\end{array}
\end{equation}
The time-stepping $\Delta{t}$ used is 0.01 seconds, from which 100 snapshots are drawn in each simulations and 40 simulations used in training. Reynolds numbers varies from 500 to 2000. Ten simulation sets are used for validation. All the data is scaled between one and zero to enable better learning by the neural network.
\paragraph{Data-driven method}
\label{sec:lid_data}
\hspace{0.5 cm} Data is used to train the model with a procedure similar to the physics-constrained case of the Burgers equation explained in section~\ref{subsec: Train physics Burgers}. The training loop schematic is given in figure~\ref{Train lid PC}, where instead of obtaining the target from governing equation, we get it from the existing data. The network trained with data accurately captured the dynamics of a lid-driven cavity for both the fields, i.e., stream function and vorticity. The stream function prediction for different time steps is shown in figure~\ref{lid_data_results}.

Here, evolution of the stream function field was fast in the initial time steps, and the network could not capture it adequately if trained for larger time steps; hence smaller time step was used.

\paragraph{Physics-constrained method}
\hspace{0.5 cm} The training using the physics-constrained model is shown in figure~\ref{Train lid PC}. Where the time steps to backpropagate are increased gradually after every 100 epoch. The approach is similar to that described in algorithm~\ref{alg: train_PC}. 
\begin {equation}
\begin{array}{l}
\label{eq: RHS spatial lid}
F=-\frac{\partial \psi}{\partial y} \frac{\partial \omega}{\partial x}+\frac{\partial \psi}{\partial x} \frac{\partial \omega}{\partial y}+\nu\left(\frac{\partial^{2} \omega}{\partial x^{2}}+\frac{\partial^{2} \omega}{\partial y^{2}}\right),\\
\end{array}
 \end{equation}
 \begin {equation}
 \label{eq: crank lid}
\begin{array}{l}
w^{*}=w^{n}+0.5 \times \Delta{t}({F^{n}+F^{n+1}}),\\
\end{array}
 \end{equation}
 \begin {equation}
\begin{array}{l}
\label{eq: psi star}
\psi^{*}=\frac{1}{4}\left[h^{2} \times {\omega_{i, j}}^{*}+{\psi^{n+1}_{i+1, j}}+{\psi^{n+1}_{i-1, j}}+{\psi^{n+1}_{i, j+1}}+{\psi^{n+1}_{i, j-1}}\right].
\end{array}
 \end{equation}
The loss function is calculated using the governing equations~\ref{eq:lid_laplace}and~\ref{eqn: vort lid} . The spatial derivatives in equation~\ref{eq: RHS spatial lid} are calculated using the previous time step, and vorticity $\omega^{*}$ at the next time step is calculated using the Crank-Nicholson scheme as shown in equation~\ref{eq: crank lid}. Stream function $\psi^{*}$ calculated from the Laplace equation~\ref{eq:lid_laplace} and its discretization shown in equation~\ref{eq: psi star} is then used to formulate the loss function  $\mathcal{L}$ calculated using mean square error given as
 \begin{equation}
 \begin{array}{l}
     \mathcal{L}_{1}: MSE(w^{n+1},w^{*}); \mathcal{L}_{2}: MSE(\psi^{n+1},\psi^{*}),\\
      \mathcal{L}=\mathcal{L}_{1}+\mathcal{L}_{2}. 
 \end{array}
 \end{equation}
 The loss function above and training using physics-constrained was able to capture the flow properties at the initial time steps but failed to capture the evolution of dynamics in time.
 


\subsubsection{Periodic vorticity formulation}
\label{sec: vort_data}
\indent For the case of lid-driven cavity, the gradients are not well defined in the corners of the cavity. This might be be the culprit for the difficulties involved in the training of the network using physics-constrained. In another attempt to solve unsteady Navier-Stokes using physics-constrained without any data, a formulation with periodic boundary conditions is tried which we call as periodic vorticity formulation. The formulation was introduced in the paper by ~\citet{Li2020Fourier}. The governing equations for the said formulations are as follows
\begin {equation}
\begin{array}{l}
\label{eq:laplace}
\begin{aligned}
\partial_{t} w(x, t)+u(x, t) \cdot \nabla w(x, t) &=\nu \Delta w(x, t)+f(x), & & x \in(0,1)^{2}, t \in(0, T] \\
\nabla \cdot u(x, t) &=0, & & x \in(0,1)^{2}, t \in[0, T] \\
w(x, 0) &=w_{0}(x), & & x \in(0,1)^{2}\\

\end{aligned}
\end{array}
 \end{equation}
 where, the forcing term $f$ is given by,
\begin{equation}
 \begin{array}{l}
f(x)=0.1\left(\sin \left(2 \pi\left(x_{1}+x_{2}\right)\right)+
\cos \left(2 \pi\left(x_{1}+x_{2}\right)\right)\right).
\end{array}
\end{equation}
\paragraph{Training using data}

\begin{figure}[H]
		\centering
		\includegraphics[width=1\linewidth]{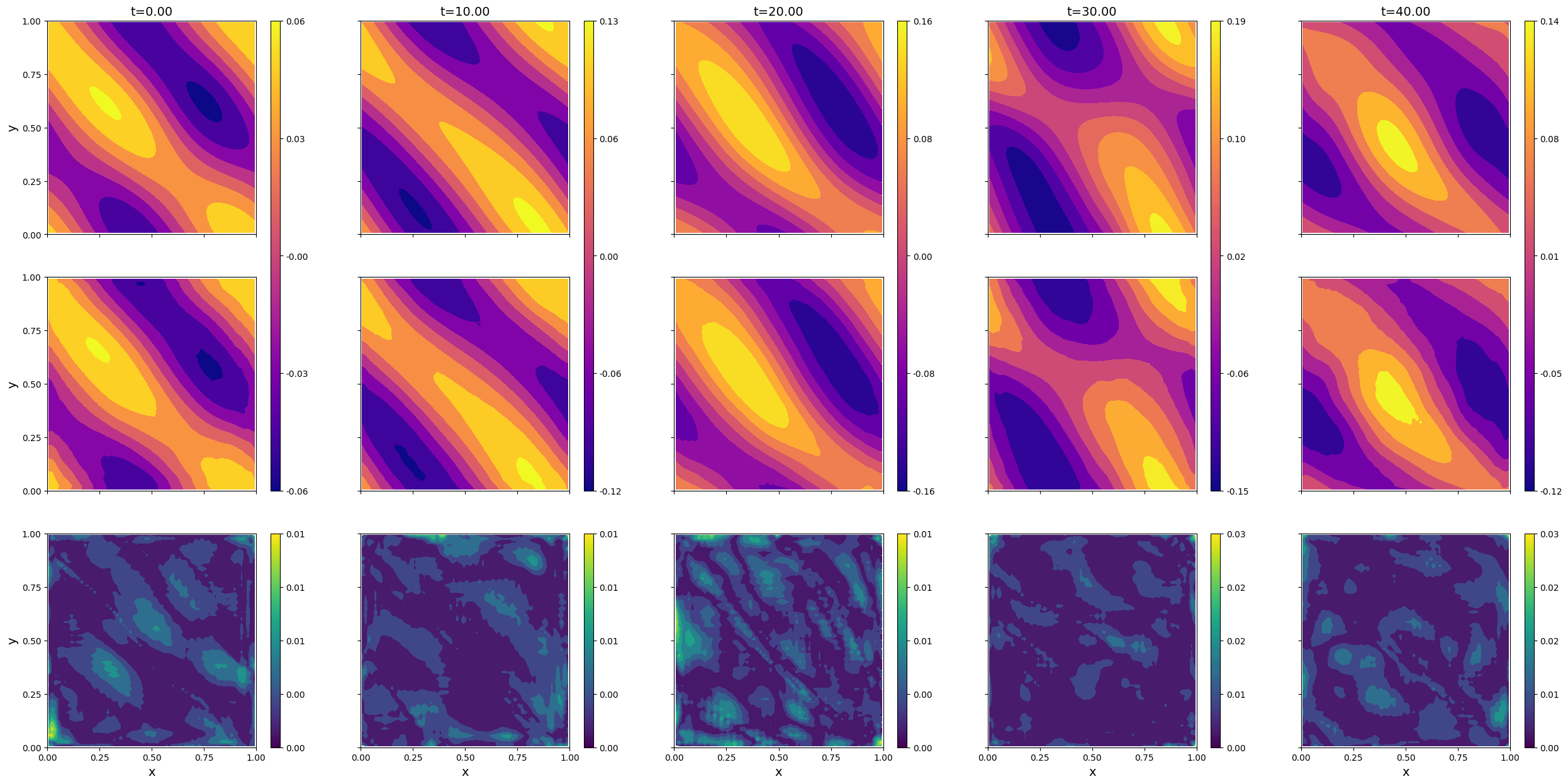}
	\caption{Prediction of the evolution of $\omega$ field in time when trained using data-driven method. (Top to bottom) finite difference target solution, AE-ConvLSTM prediction, $L_1$ error.}
	\label{vort_data_results}
\end{figure}

\hspace{0.5 cm} The data-driven case for the formulation involving all three fields and 4000 simulations for 50 time steps yielded excellent results as shown in vorticity plots in figure~\ref{vort_data_results}. The training of the network is similar to that of the lid-driven case as explained in section~\ref{sec:lid_data}. 
	
\begin{figure}[H]
		\centering
			\includegraphics[width=1\linewidth]{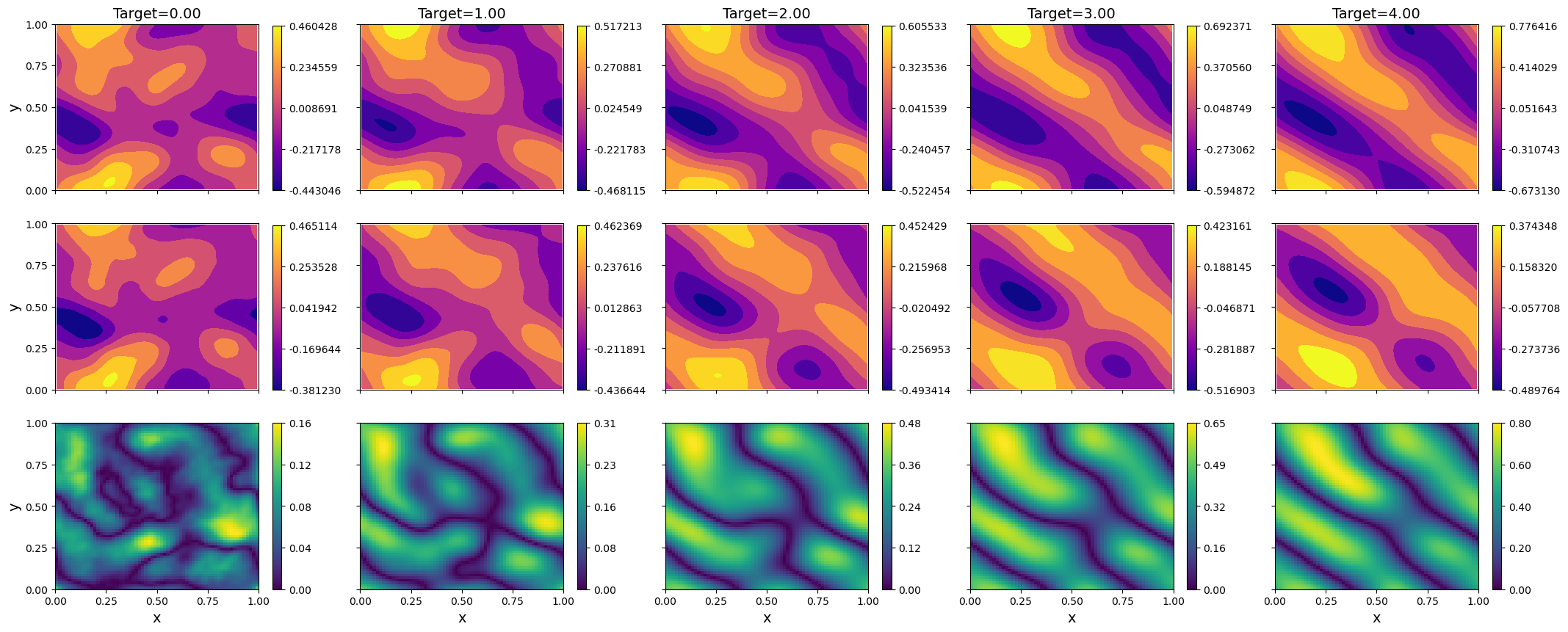}
    	\caption{First 5 time step prediction of the evolution of $\omega$ field when trained using physics-constrained method. (Top to bottom) finite difference target solution, AE-ConvLSTM prediction, $L_1$ error.}
    	\label{vort_PC}
	\end{figure}

\paragraph{Physics-constrained method}
\hspace{0.5 cm} Attempts for physics-constrained case were done using the velocity-vorticity formulation discretized as

\begin {equation}
\begin{array}{l}
F = -u \frac{\partial \omega}{\partial x}-v\frac{\partial \omega}{\partial y}+\nu\left(\frac{\partial^{2} \omega}{\partial x^{2}}+\frac{\partial^{2} \omega}{\partial y^{2}}\right)+f(x),\\

w^{*}=w^{n}+0.5 \times \Delta{t}({F^{n}+F^{n+1}}),\\

Residual=\left(\frac{\partial u}{\partial x}+\frac{\partial v}{\partial y}\right).
\end{array}
 \end{equation}
where loss is calculated using the following formulation $\mathcal{L}: MSE(w^{n+1},w^{*})+Residual$.\\
\indent The formulation was able to capture first few time steps for the vorticity but failed to capture its time evolution thereafter. The snapshots of the prediction of vorticity field $\omega$ when trained with physics-constrained method is shown in figures~\ref{vort_PC}.


\section{Conclusion}~\label{sec: conclusion}
\indent An AE-ConvLSTM network is introduced, which predicts better than a ConvLSTM network~\citet{Shi2015CLSTM} when trained on data. It also predicts a longer sequence by enabling multiple passes through the network by carrying hidden states from one encoder to another. Hence, AE-ConvLSTM is an excellent network to use when there is a need to predict the time series of a dynamical system. Its sequence to sequence nature helps to alleviate the problem of vanishing gradients during training as fewer network graphs are needed to be stored for back-propagation.\\
\indent Further, AE-ConvLSTM was tested to predict the 2-D viscous Burgers equation, and the network could predict 100-time steps when trained on data and was later trained without data using governing equations (i.e., physics-constrained). The training with physics-constrained enabled the prediction of the viscous Burgers problem for 100 time-steps when trained until 35-time steps.\\
\indent AE-ConvLSTM's ability to predict the evolution of unsteady Navier-Stokes solution was tested by training the network with data and by using governing equations for the case of the lid-driven cavity and a periodic vorticity formulation. It is observed that the AE-ConvLSTM network's performance was excellent in predicting the flow evolution when trained on data. Still, the network performed poorly when trained on governing equations without data (physics-constrained). The reason could be the occurrence of local minima while trying to optimize the loss function when the solution is evolving through time. For N-S, pressure and velocity evolve at different rates in time, and AE-ConvLSTM fails to predict it. Time-evolving multi-scaled coupled fields could be a pain point for physics-constraint training in general, and future work is being carried out to alleviate this issue. 

\bibliography{Ref}
\end{document}